\documentclass{aa}  

\usepackage{graphicx}
\usepackage{txfonts}
\usepackage[T1]{fontenc}
\begin{document}

   \title{New nickel opacities and their impact on stellar models}
   \titlerunning{New nickel opacities}


   \author{A. Hui-Bon-Hoa
          \inst{1},
          J.-C. Pain\inst{2,3},
          O. Richard\inst{4}
          }

   \institute{IRAP, Universit\'e de Toulouse, CNRS, UPS, CNES, Toulouse, France\\
            \email{alain.hui-bon-hoa@irap.omp.eu}
   \and
   CEA, DAM, DIF, F-91297 Arpajon, France
   \and
   Universit\'e Paris-Saclay, CEA, Laboratoire Mati\`ere sous Conditions Extr\^emes, F-91680 Bruy\`eres-le-Ch\^atel, France
   \and
   Laboratoire Univers et Particules de Montpellier, Universit\'e de Montpellier, CNRS, Place Eug\`ene Bataillon, 34095 Montpellier,
France}

   \date{Received ; accepted }

 
  \abstract
   {The chemical element nickel is of particular interest in stellar physics. In the layers in which the Fe-peak elements dominate the mean opacity (the so-called Z-bump), Ni is the second contributor to the Rosseland opacity after iron, according to the Opacity Project data. Reliable nickel cross sections are therefore mandatory for building realistic stellar models, especially for main-sequence pulsators such as $\beta$ Cep and slowly pulsating B stars, whose oscillations are triggered by the $\kappa$-mechanism of the Fe-peak elements. Unfortunately, the Opacity Project data for Ni were extrapolated from those of Fe, and previous studies have shown that they were underestimated in comparison to detailed calculations.}
   {We investigate the impact of newly computed monochromatic cross sections on the Rosseland mean opacity of Ni and on the structure of main-sequence massive pulsators. We compare our results with the widely used Opacity Project and OPAL data.}
   {Monochromatic cross sections for Ni were obtained with the SCO-RCG code. The Toulouse-Geneva evolution code was used to build the stellar models.}
   {With the new data, the Rosseland opacities of Ni are roughly the same as those of the Opacity Project or OPAL at high temperatures ($\log\ T>6$). At lower temperatures, significant departures are observed; the ratios are up to six times higher with SCO-RCG. These discrepancies span a wider temperature range in the comparison with OPAL than in comparison with the Opacity Project. For massive star models, the results of the comparison with a structure computed with Opacity Project data show that the Rosseland mean of the global stellar mixture is only marginally altered in the Z-bump. The maximum opacity is shifted towards slightly more superficial layers. A new maximum appears in the temperature derivative of the mean opacity, and the driving of the pulsations should be affected.}
   {}


   \maketitle
%

\section{Introduction}

Opacities describe the interactions between radiation and matter. In stellar physics, they are used to evaluate the amount of energy that is carried by the radiation field. They are therefore of prime importance for building stellar models. In stellar interiors, the medium is optically thick so that the transfer equation can be expressed by an analytical function that involves the Rosseland mean opacity (RMO). In most parts of main-sequence stars, which are in the hydrogen-burning phase, the most important elements that contribute to the RMO are hydrogen and helium. One noticeable exception is the layer around $\log T=5.3,$ where the iron peak elements are the main contributors (this is often referred to as the Z-bump).\\
In addition to structural effects, the domination of the RMO by the Fe-peak elements enables them to trigger oscillations  through the $\kappa$-mechanism for stars in the upper part of the main sequence. This mechanism is invoked to explain the pulsations of the slowly pulsating B (SPB) stars \citep{Dziembowski_etal1993}, which are the coolest pulsators of the upper main sequence and exhibit high-overtone gravity modes \citep{Waelkens1991}. In turn, in the $\beta$ Cep pulsators \citep[e.g.][and references therein]{Stankov_Handler2005}, this process excites low-order $p$- and $g$-modes \citep{Dziembowski_Pamyatnykh1993}. In the overlapping zone between the instability strips of the SPB and $\beta$-Cep (around 10~$M_\odot$), some objects show oscillations of both types of stars \citep[the so-called hybrid pulsators;][and references therein]{Balona_2011}, which are also excited by the Z-bump $\kappa$-mechanism \citep{Pamyatnykh1999}. Outside the main sequence, some B-type subdwarfs (sdB, $22,000\lesssim T_\mathrm {eff}\lesssim 35,000$) also show pulsations. Their effective temperature range is similar to that of the $\beta$ Cep and the hottest SPBs.
\cite{Charpinet_etal1996,Charpinet_etal1997} and \cite{Fontaine_etal2003} showed that the driving mechanism for the $p$-mode and for the $g$-mode oscillations in sdBs, respectively, is the $\kappa$-mechanism of the Fe-peak elements.

The need for sufficiently complete and accurate opacities to calculate realistic stellar models led to the computations of theoretical opacity dataset starting in the mid-1990s. Of the different groups involved in this effort, the Opacity Project (OP) is the only one to have made the monochromatic cross sections for each chemical element publicly available \citep{Seaton2005}. The mean opacities can then be computed consistently with respect to the local mixture in each part of the star. The monochromatic opacities of Cr, Mn, and Ni were not computed in detail, however, and come from an extrapolation of the data for iron \citep{Seaton_etal1994}. \cite{Turck_Chieze_etal2016} showed that this method could lead to discrepant values of the RMO especially for Ni compared to detailed computations, so that a caveat was attached to the reliability of the RMO computed with the OP opacities in the Z-bump. In addition, nickel is of prime importance in the excitation of the pulsations, especially for high-overtone $g$-modes in upper main-sequence stars \citep[e.g.][]{Daszynska-Daszkiewicz_etal2017a} or in sdBs \citep{Jeffery_Saio2006}.\\
Monochromatic data have also been produced by the OPAL group \citep{Iglesias_Rogers1996} and were used in the Montr\'eal \citep{Turcotte_etal1998a} and Montr\'eal/Montpellier \citep{Deal_etal2021} stellar evolution codes, but they are not publicly available. \cite{Hui-Bon-Hoa_Vauclair2018b} found that the radiative accelerations for nickel in the Z-bump are of similar strength between OP and OPAL, which poses the question whether the OPAL opacities might also be underestimated there. \\
Accurate data for this element are therefore needed to build realistic models. Rosseland means with detailed calculation of the Ni opacities are available in the OPLIB database \citep{Colgan_etal2016}, which is suitable for models with homogeneous content in their envelope. Models in which atomic diffusion is considered exhibit inhomogeneous chemical compositions \citep{Hui-Bon-Hoa_Vauclair2018b,Hui-Bon-Hoa_Vauclair2018} and require monochromatic cross sections in order to consider the feedback of the local abundances on the opacities. We therefore present new monochromatic data computed with the SCO-RCG code \citep{Pain_Gilleron2015}.

After a description of the SCO-RCG code in Sect.~\ref{sco_rcg}, we present the new data and compare their RMOs with those of OP and OPAL (Sect.~\ref{data}). We then explain how we processed our data to use them in stellar evolution codes (Sect.~\ref{implementation}). Finally, we discuss the impact of the new data on a stellar structure computed with the Toulouse-Geneva evolution code (TGEC) in Sect.~\ref{stellarModel}.

\section{SCO-RCG code}\label{sco_rcg}

In this section, we outline the main features of the SCO-RCG code, which is described in detail in \cite{Pain_Gilleron2015}. SCO-RCG is an opacity computation code consisting of two parts: the building of the atomic structure that consistently takes plasma effects into account (and assumes local thermodynamical equilibrium) is devoted to the super-configuration code for opacity (SCO) part \citep{Blenski_etal2000}, thus avoiding the isolated-atom approximation. A self-consistent calculation is performed for each super-configuration, which in this way has its own potential and set of wave functions. That density effects on the wave functions are considered is one of the strengths of SCO-RCG compared to other codes. 
The data required to calculate the detailed transition arrays (i.e. direct and exchange Slater, spin-orbit, and dipolar integrals) are calculated by the SCO part within a plasma model accounting for the density and screening effects on the wave functions, and given as inputs to the RCG part \citep{Cowan1981}, performing the detailed-line computation of the transition array (detailed-line accounting, DLA). The selection of transition arrays for which a detailed line-by-line treatment is possible and relevant is made according to some user-defined criteria involving the mean energy spacing between neighbouring lines and the mean line width in the transition array. DLA calculations are performed only for pairs of configurations giving rise to fewer than 800,000 lines (the maximum size of a $J$-block inside a configuration is 4000). In other cases, transition arrays are represented statistically by Gaussian profiles in the unresolved transition arrays \citep[UTA;][]{Bauche1979} or spin-orbit split arrays \citep[SOSA;][]{Bauche1985} formalisms. If the Rydberg supershell contains at least one electron, then transitions starting from the super-configuration are treated within the super-transition array model \citep[STA;][]{BarShalom1989}.\\
This hybrid design allows us to consider many highly excited states that can have a significant contribution to the opacity in spite of their low probabilities. The number of detailed calculations in SCO-RCG is now largely dominant, and subsequently the computed spectrum is less sensitive to the modelling of the remaining statistical contributions (UTA, SOSA, and STA).

\section{Data for nickel}\label{data}

\subsection{Rosseland mean opacities}\label{RMO}

\begin{figure}[ht]
   \centering
   \includegraphics[width=.5\textwidth]{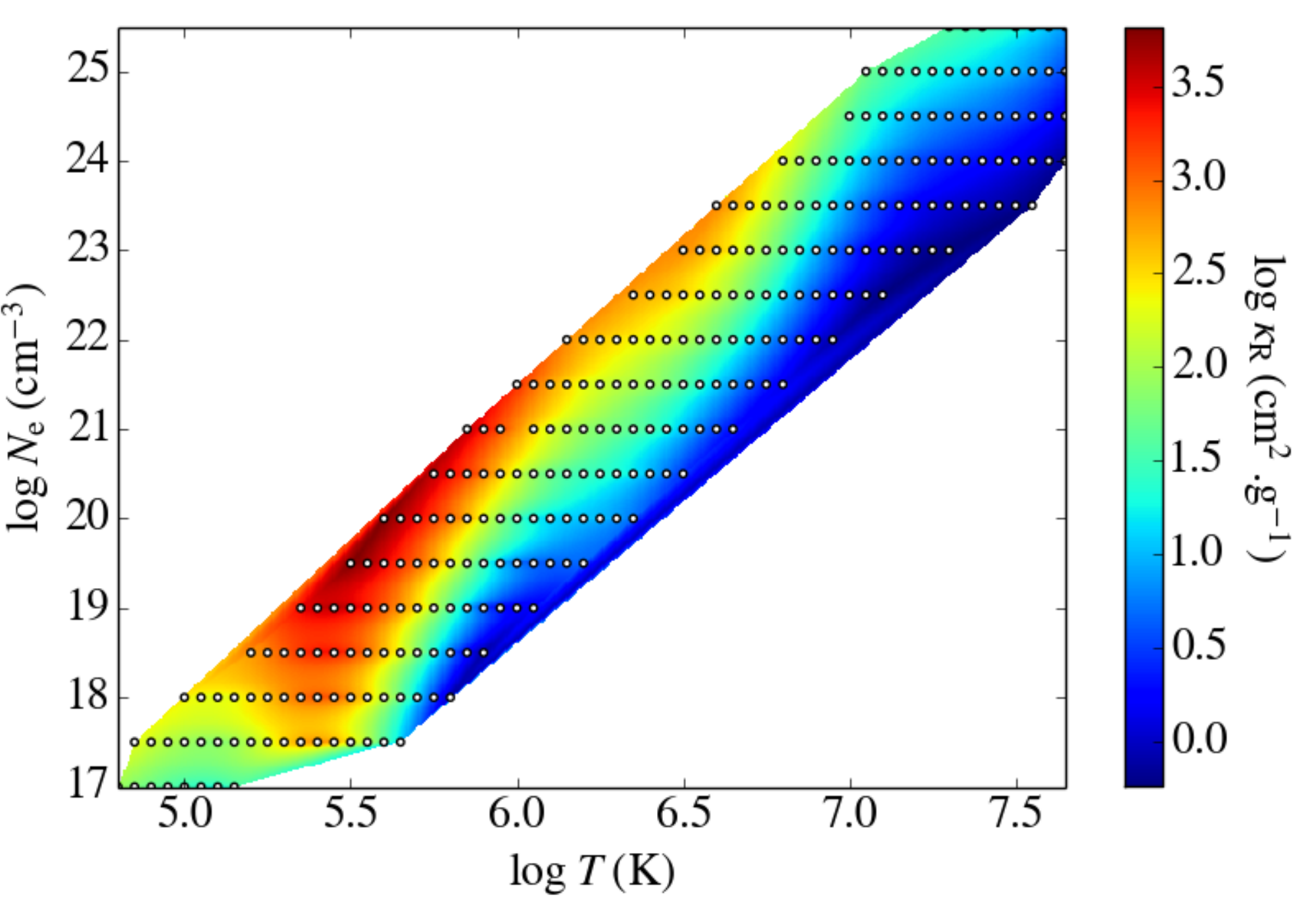}
      \caption{Rosseland mean opacities $\kappa_{\mathrm R}$  of nickel computed with SCO-RCG in the ($T$, $N_{\mathrm e}$) plane. The dots represent the grid points where the computations are made. The highest values are met in the area around $\log T=5.4$, i.e., in the so-called Z-bump.
              }
         \label{figure:RMO}
\end{figure}

Among the different types of main-sequence stars, seismic observations are the most stringent for the hybrid pulsators, which simultaneously exhibit acoustic and gravity waves ($p$- and $g$-modes, respectively). Therefore, we computed the cross sections for Ni in the temperature-electron density ($T$, $N_{\mathrm e}$) domain involved in the main-sequence evolution of such a star. In practice, we chose a mass of 9.5~$M_\odot$, which is the mass of the well-studied hybrid pulsator $\nu$ Eri \citep[e.g.][]{Handler_etal2004,Pamyatnykh_etal2004,Daszynska-Daszkiewicz_etal2017a}. This domain appears as a strip in this plane. The new data were calculated at the same 371 grid points as the OP cross sections since they are meant to be combined together to compute the RMO for the actual chemical mixture of a stellar model. The computations were successful at the 274 grid points shown in Fig.~\ref{figure:RMO}. They failed for 97 points, especially at low temperatures and electron densities $(\log T<4.7, \log N_\mathrm{e}<17)$, and for $\log N_\mathrm{e}=17,$ only the points with $\log T \in [4.8 ;5.15]$ could be computed. For all these points, the Fermi-Dirac distribution function tends to a Heaviside function due to the low densities, leading to numerical issues. In addition, five other points ($(\log T,\log N_\mathrm{e})$=(6,21), (6.75,24), (6.9,24.5), (6.95,24.5), and (7.45,25.5)), scattered among the successful calculations, could not be computed for a reason still under investigation, yielding the gaps in Fig.~\ref{figure:RMO}. We show in the next subsection, however, that the effect on the stellar structure should be small. The highest values are in the Z-bump, around $\log T=5.4$.\\
We recall that the Rosseland opacity $\kappa_\mathrm{R}$ is a harmonic mean of the monochromatic opacities $\kappa_\nu$ weighted by the temperature derivative of the Planck function $\frac{dB_\nu}{dT}$,
\begin{equation}
\frac{1}{\kappa_\mathrm{R}}=\frac{\int_{0}^{\infty} \frac{\frac{dB_\nu}{dT}}{\kappa_\nu} \mathrm{d}\nu}{\int_{0}^{\infty} \frac{dB_\nu}{dT}  \mathrm{d}\nu}
,\end{equation}
which can also be expressed as
\begin{equation}\label{eq:RMO}
\frac{1}{\kappa_\mathrm{R}}=\int_{0}^{\infty} \frac{F(u)}{\kappa(u)} \mathrm{d}u
,\end{equation}
where $F(u)$ is the normalised temperature derivative of the Planck function in terms of $u=\frac{h\nu}{k\mathrm{T}}$ \citep[the expression of $F(u)$ is given by Eq.~8 of][]{Badnell_etal2005}. In this study, all the values of Rosseland means are computed with a trapezoidal rule.

\begin{figure*}
   \centering
   \includegraphics[width=\textwidth]{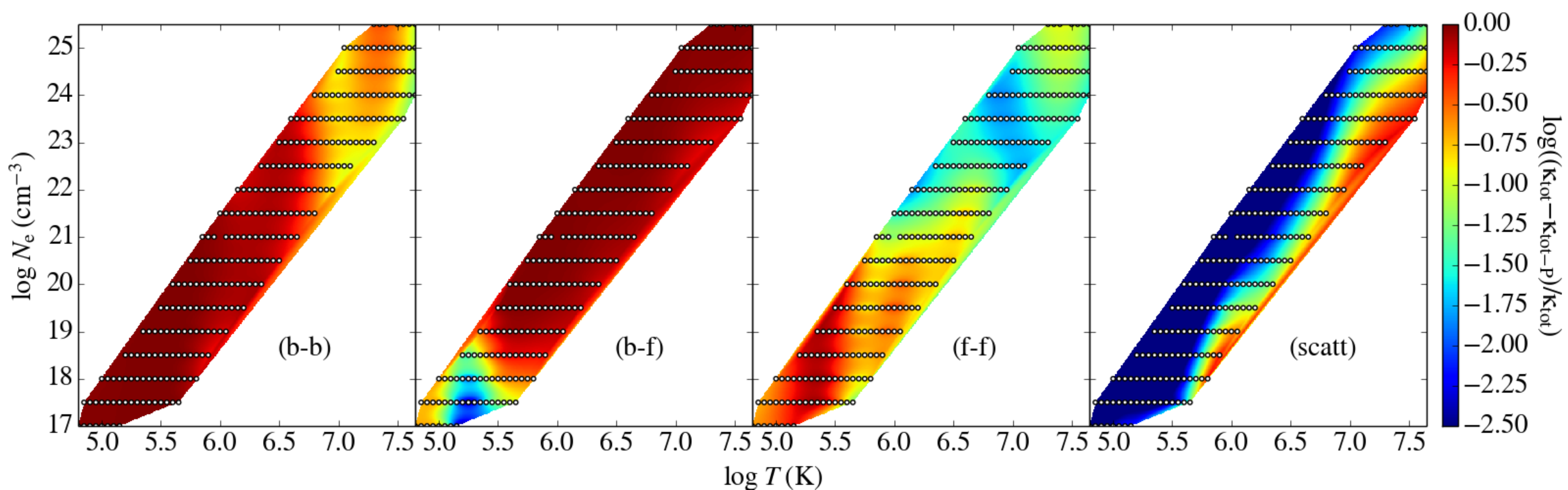}
      \caption{Logarithm of the contributions to the Rosseland mean opacities (see text) in the ($T$, $N_{\mathrm e}$) plane, plotted according to the same scale. Its lower limit is set to -2.5, and the smallest scattering contributions are upper limits. $\kappa_{\mathrm {tot}}$ is the RMO considering all the absorption processes, $\kappa_{\mathrm {tot-P}}$ is the RMO without process P, P being one of the opacity sources (lines b-b, photoionisation b-f, reverse bremsstrahlung f-f, or scattering, scatt).              }
         \label{figure:contribRMO}
\end{figure*}

The total monochromatic opacity $\kappa_\nu$ or $\kappa(u)$ is the sum of the cross sections of four physical processes: absorption through spectral lines (bound-bound transitions), photoionisation (bound-free transitions), reverse bremsstrahlung (free-free interactions), and electron scattering. Figure~\ref{figure:contribRMO} shows the contribution of each process to the RMO. A contribution is defined as $(\kappa_{\mathrm {tot}}-\kappa_{\mathrm {tot-P}})/\kappa_{\mathrm {tot}}$ 
, where $\kappa_{\mathrm {tot}}$ is the Rosseland mean considering all the absorption processes and $\kappa_{\mathrm {tot-P}}$ the RMO with process P ignored. If process P has a significant role in the RMO, then $\kappa_{\mathrm {tot-P}}$ is small compared to $\kappa_{\mathrm {tot}}$ and its contribution tends to 1. Conversely, if its role is negligible, $\kappa_{\mathrm {tot-P}}$ is close to $\kappa_{\mathrm {tot}}$ and the contribution is then close to zero. We conclude from Fig.~\ref{figure:contribRMO} that the RMOs are mostly due to line absorptions and to photoionisation, the free-free process contributing marginally at low temperatures. Bound-bound absorption is no longer a dominant process above $\log T\simeq6.7$, where the number of electrons per atom is higher than 24, in other words, fewer than four electrons are left on average. Scattering is the less important process everywhere in the computed domain.

\begin{figure}
   \centering
   \includegraphics[width=.5\textwidth]{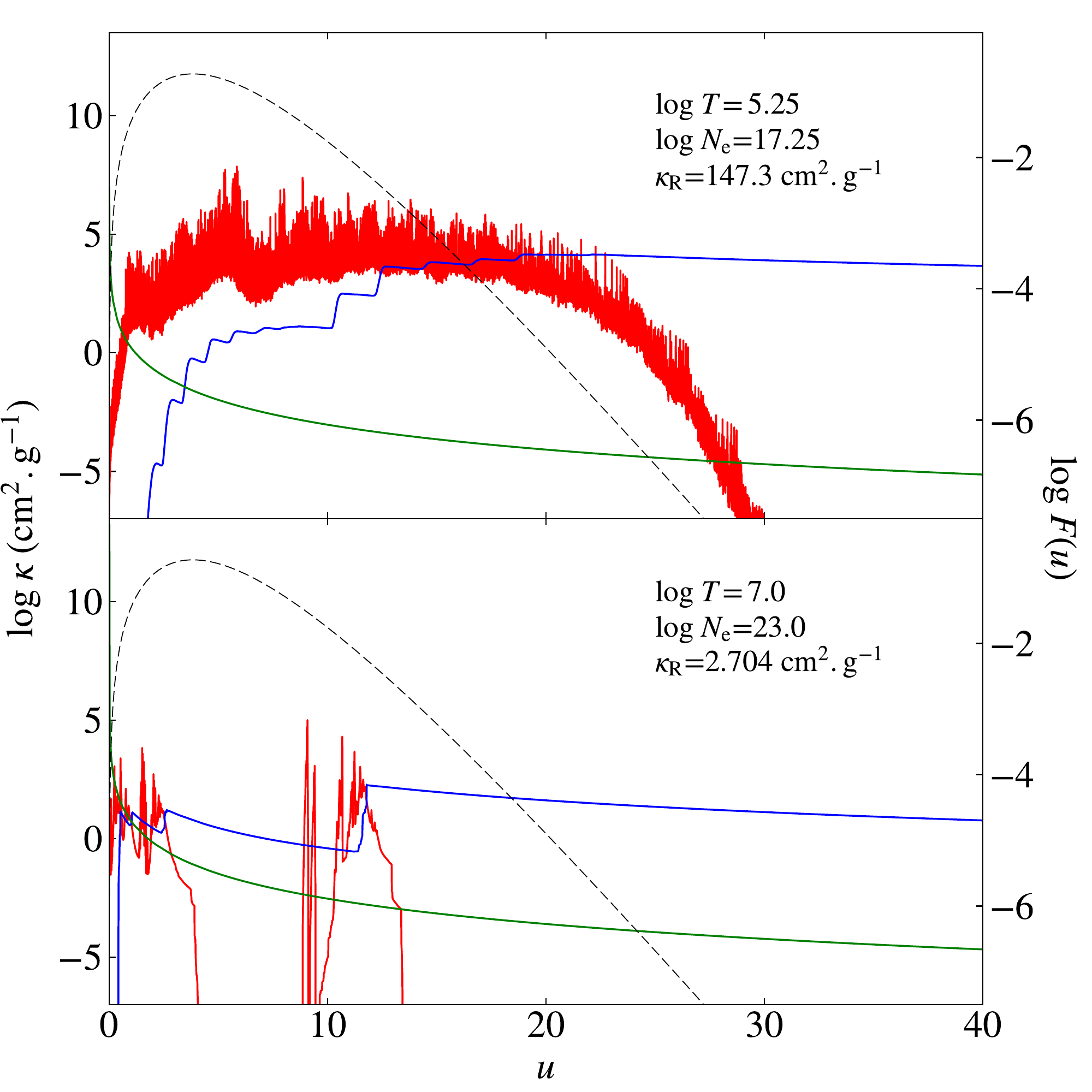}
      \caption{Spectra with the logarithm of the spectral lines (red lines), photoionisation (blue lines), and reverse-bremsstrahlung (green lines) cross sections (left scale) vs. photon energy expressed in $u=\frac{h\nu}{k\mathrm{T}}$ for two different temperature-electron density sets. The conditions for which they have been computed and the resulting Rosseland mean are detailed in each panel. The dashed lines denote the normalised temperature derivative of the Planck function in logarithm, $\log\ F(u)$ (right scale). 
              }
         \label{figure:spectra}
\end{figure}

Figure~\ref{figure:spectra} shows the spectra of the different absorption processes for two sets of ($T$, $N_{\mathrm e}$): the set shown in the upper panel is in the Z-bump, more specifically, in a region in which line absorption dominates the RMO and where the contribution of photoionisation is the smallest. The lower panel refers to conditions where photoionisation is the main absorption process and where the bound-bound opacity is close to its lowest contribution. The effect of the weighting by the temperature derivative of the Planck function is evident: any absorption process that has significant cross sections around the maximum of $F(u)$ will be a main contributor to the RMO. Scattering is not shown not only because of its negligible contribution, but mostly because it is not considered in the rest of this work. Its actual effect is to be computed after the mixture of the medium defined, so that the number of free electrons per atom and the ionisation fractions for the different elements in the stellar conditions are known.

\subsection{Comparison with other data}\label{compareOP}

\subsubsection{Opacity Project}
\begin{figure}
   \centering
   \includegraphics[width=.5\textwidth]{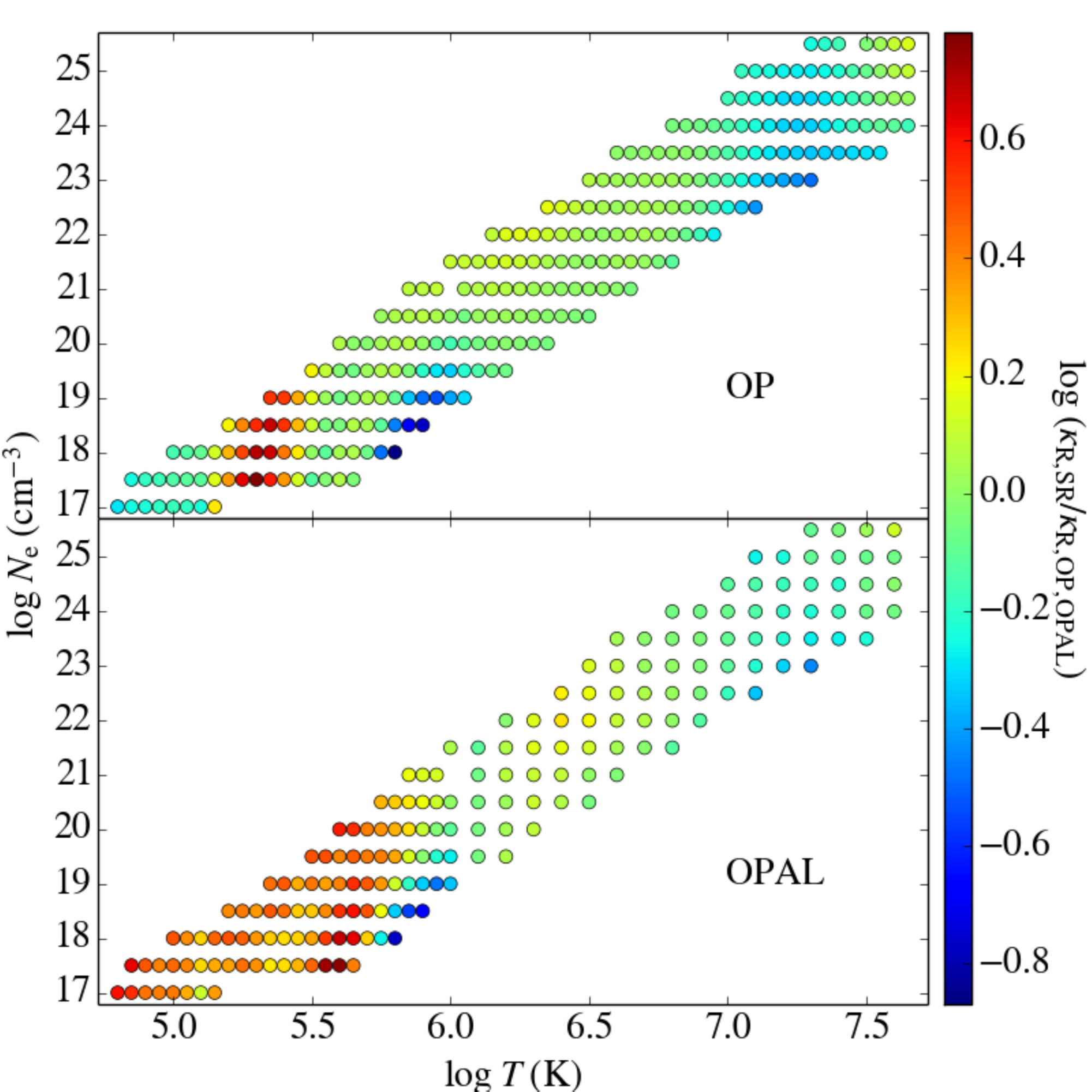}
      \caption{Ratios of the SCO-RCG Rosseland means and those of OP or OPAL in the ($T$, $N_{\mathrm e}$) plane. $\kappa_\mathrm{R, SR}$ is the RMO from the new data and $\kappa_\mathrm{R, OP, OPAL}$ that of the comparison dataset. {\it Top panel:} Comparison with OP. The SCO-RCG RMOs can be up to six times larger than those of OP in a narrow band in the Z-bump, around $\log T=5.3$. {\it Bottom panel:} Same plot with OPAL. The OPAL temperature step is larger for $\log\ T>6$. The area in which the SCO-RCG RMOs are higher than those of OPAL is larger than for OP.
              }
         \label{figure:RMOdifference}
\end{figure}

The Ni cross sections of OP come from an extrapolation of those of iron \citep{Seaton_etal1994}. \cite{Turck_Chieze_etal2016} showed that the RMOs computed with these values could be very different from that of detailed calculations. The upper panel of Fig.~\ref{figure:RMOdifference} presents the logarithms of the ratios of the RMOs computed with our new data ($\kappa_\mathrm{R, SR}$) and those from OP ($\kappa_\mathrm{R, OP}$) in the ($T$, $N_{\mathrm e}$) plane. The values of $\kappa_\mathrm{R, OP}$ were computed from the monochromatic cross sections using a trapezoidal scheme.\\
The ratios of the SCO-RCG and the OP RMOs can be as high as 6 in the Z-bump, that is, of the same order of magnitude as the value stated by \cite{Turck_Chieze_etal2016}. Outside the Z-bump, the RMOs are roughly similar, except for several grid points at which the RMO from the new data can be as low as one-tenth of that of OP. Because Ni is no longer a main contributor to the RMO outside the Z-bump, however, we expect that the effect on the stellar structure should be small compared to the use of OP data. The case of layers whose temperature is in the Z-bump, but whose electronic density is lower than the lower limit of the computed domain is more questionable. Further calculations towards lower densities are needed to determine to which extent the RMOs are underestimated with the OP cross sections in these conditions.

\subsubsection{OPAL}

Although the cross sections of iron-peak elements were recently updated \citep{Iglesias_2015}, they are not available to us. We therefore show the results from previous computations here \citep{Iglesias_Rogers1996}. Nevertheless, the impact of the new OPAL calculations should be negligible according to Fig.~1 of \cite{Iglesias_2015}. The OPAL opacities were computed in detail for all the elements of their set, at grid points defined by the temperature and $R=\rho/T6,$ where $\rho$ is the density and $T6$ is the temperature divided by $10^6$. The temperature values are the same as those of OP up to $\log\ T=6$, above which the step is twice that of OP in $\log\ T$. Since the spacing is constant in $\log\ R$, the electronic densities at the OPAL grid points are different from those of our grid. We obtained the Rosseland means at our $N_\mathrm e$ values by interpolating the OPAL RMOs using a cubic spline. To be consistent with the comparison with OP, the RMOs were also computed from the monochromatic cross sections.\\
The logarithms of the ratios of the SCO-RCG RMOs and those from OPAL ($\kappa_\mathrm{R, OPAL}$) are plotted in the lower panel of Figure~\ref{figure:RMOdifference}. As for OP, the new data yield higher Rosseland means in the Z-bump, but the area covered by these enhancements is larger. At the same location as in the comparison with OP, several values are lower than those from OPAL, with likely little influence on the stellar structure as they lie outside the Z-bump. These discrepancies could arise from the number of atomic configurations involved in the calculations. A more thorough investigation of the impact of the difference between SCO-RCG and OPAL would require computing the SCO-RCG cross sections at the same grid points as OPAL.

\section{Implementation in stellar codes}\label{implementation}

\begin{figure*}
   \centering
   \includegraphics[width=\textwidth]{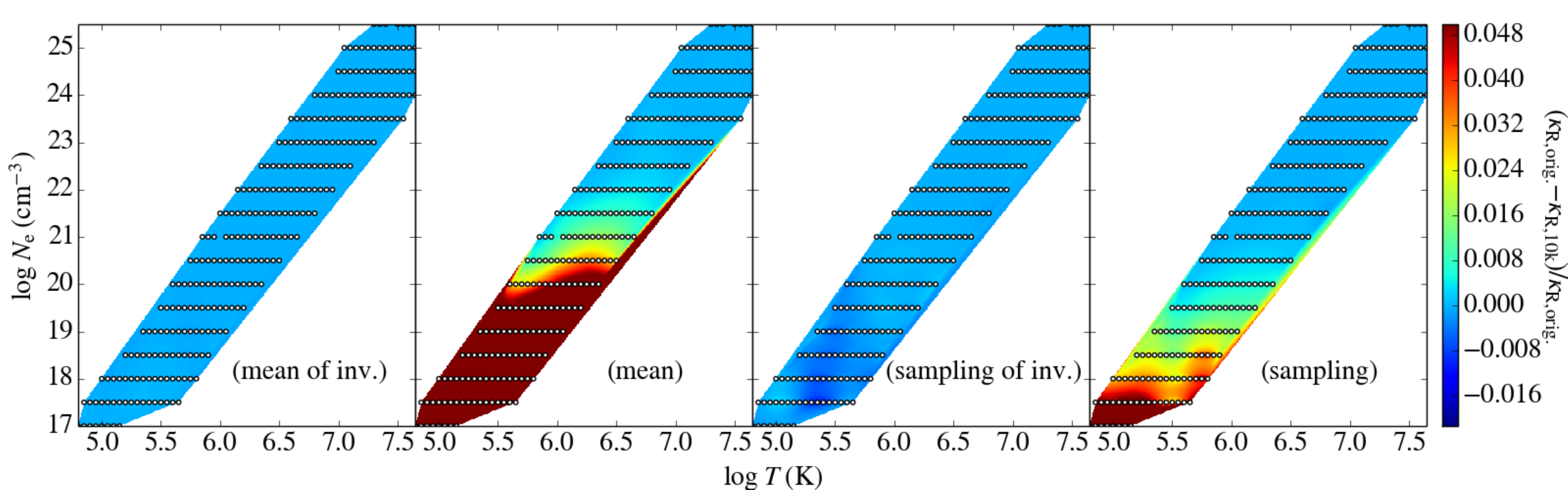}
      \caption{Relative difference between the original SCO-RCG RMO ($\kappa_\mathrm{R, orig.}$) and the calculation with the OP frequency mesh ($\kappa_\mathrm{R, 10k}$) in the ($T$, $N_{\mathrm e}$) plane, plotted according to the same scale. Its upper limit is set to 5~\%. The dark red parts are to be considered as lower limits. From left to right, the panels show the results with the mean of the inverses of the cross sections, the mean of the cross sections, the sampling of the inverses, and the sampling of the cross sections.
              }
         \label{figure:reduc10000}
\end{figure*}

For stellar evolution codes to benefit from the new data, we aim to use them in lieu of the existent data (OP or OPAL). We also wished to reach this goal without changing the routine that computes the Rosseland mean, so that we have to replace the existent data by those obtained with the new calculations whenever available. The physical quantities to change are obviously the cross sections, but also the mean number of electrons per atom and the ionisation fractions for the various elements that are needed to compute the scattering. Since the SCO-RCG cross sections are computed at present on the same grid points as the OP data, we focused on introducing the new data in the OP files. Further efforts are needed to merge the new opacities with OPAL data.\\
The cross sections of SCO-RCG were computed for a photon energy grid comprising 299,999 points equally spaced in $u$ in the interval between 0 and 40. In turn, the OP mesh contained 10,000 points with a spacing keeping $F(u)\Delta u$ roughly constant, with $\Delta u$ the step between consecutive grid points. $u$ itself is between 0 and 20. We tested two methods to reduce the number of mesh points, one with a sampling of the SCO-RCG data on the OP grid points, the other by considering the average between the middle of the two intervals surrounding each OP grid point, which could provide a more representative value where the spectra have complex features. Furthermore, Eq.~\ref{eq:RMO} suggests that the inverses of the cross sections shoud be used to calculate the RMOs, rather than the cross sections themselves. In addition to these two calculations, we applied both methods to the cross sections themselves to determine the relevance of using their inverses. We defined the relative change of the RMO as 
$(\kappa_\mathrm{R,orig.}-\kappa_\mathrm{R,10k})/\kappa_\mathrm{R,orig.}$ where $\kappa_\mathrm{R, orig.}$ is the RMO obtained with the original energy grid and $\kappa_\mathrm{R, 10k}$ with that of OP. These differences are plotted in the ($T$, $N_{\mathrm e}$) plane in Fig.~\ref{figure:reduc10000} for each combination. The mean of the inverses of the cross sections definitely provides the best results, whereas the mean of the cross sections yields the largest differences. We note that these cross sections do not include the contribution of electron scattering.\\
The last step of the implementation consists of replacing the original OP cross sections, number of electrons per atoms, and ionisation fractions in the OP files. For this, the new cross sections were converted into atomic units, which is the unity used in the OP files.


\section{Impact on stellar models}\label{stellarModel}

TGEC was used as testbed. TGEC is a 1D stellar evolution code \citep{Hui-Bon-Hoa2008} that implements the latest physics. The RMOs were computed on-the-fly with the method described in \cite{Hui-Bon-Hoa2021}. This ensures full consistency between the mean opacity and the chemical composition in any layer of the model. We used the monochromatic opacities from the Opacity Project OPCD v.3.3 data \citep{Seaton2005} for all the chemicals but nickel, for which the new data were considered whenever available. The computations were performed for a $9.5~M_{\odot}$ model without atomic diffusion, so that the chemical composition is homogeneous outside the stellar core, where nucleosynthesis takes place. We used the \cite{Grevesse_Noels1993} metal content, as in \cite{Hui-Bon-Hoa_Vauclair2018b,Hui-Bon-Hoa_Vauclair2018}. Using other solar compositions such as that of \cite{Asplund_etal2009} should not affect the bulk of our results because the ratios of the different Fe-peak elements and H differ from those of \cite{Grevesse_Noels1993} by less than 0.1~dex.

\begin{figure}
   \centering
   \includegraphics[width=.5\textwidth]{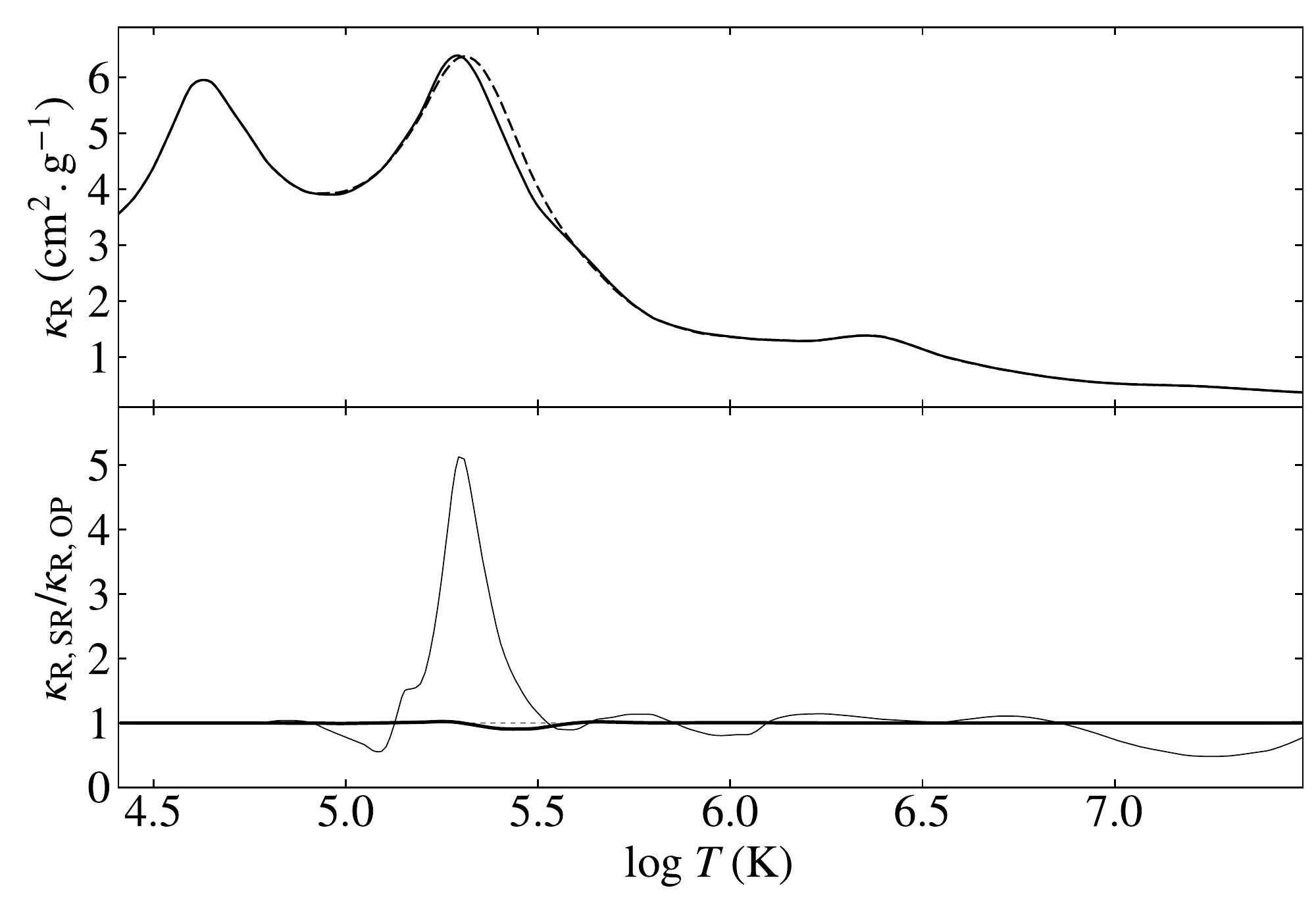}
      \caption{Comparison of the mean opacities of a stellar model computed with the new Ni data and one with the OP data. {\it Upper panel}:  Rosseland mean ($\kappa_\mathrm{R}$) of the stellar mixture with the new Ni data (solid line) and that with OP (dashed line) vs. $\log\ T$. {\it Lower panel}: Ratio of the SCO-RCG ($\kappa_\mathrm{R, SR}$) and OP ($\kappa_\mathrm{R, OP}$) Rosseland means vs. $\log\ T$ for the stellar mixture (thick line) and for Ni alone (thin line). The dotted line represents a ratio of one.
              }
         \label{figure:RMOratio}
\end{figure}

We first compared the impact of the new Ni opacities on the Rosseland mean of the whole chemical mixture of the star. The upper panel of Fig.~\ref{figure:RMOratio} shows the RMOs of a stellar structure computed with the new Ni data along with those of a structure calculated with the original OP data as a function of temperature. The new data shift the maximum opacity of the Z-bump towards a slightly lower temperature, and the width of the peak is reduced. In his comparison of OP and OPAL, \cite{Iglesias_2015} suspected that the greater width of the Z-bump obtained with the OP cross sections could be due to Ni, which is consistent with our results. The value of the maximum is marginally enhanced (by about $2\times 10^{-3}$). No significant difference can be seen outside the Z-bump, as expected. With the same abscissae, we plot in the lower panel the ratio of the RMOs for the whole stellar mixture computed with the new data and those from the OP data (thick line). The same ratio for Ni alone is also shown (thin line). The dip of the global RMO ratio around $\log\ T= 5.5$ accounts for the change in shape of the Z-bump. The ratio for nickel alone has a maximum close to 5 around $\log\ T=5.3$, to be related to the high ratios present in Fig.~\ref{figure:RMOdifference} at similar temperatures. No clear correlation appears between the RMO ratio for Ni and the global ratio, however. In particular, in some layers of the Z-bump around $\log\ T=5.4$, the new Ni RMO can even be higher along with a lower global RMO.

\begin{figure}
   \centering
   \includegraphics[width=.5\textwidth]{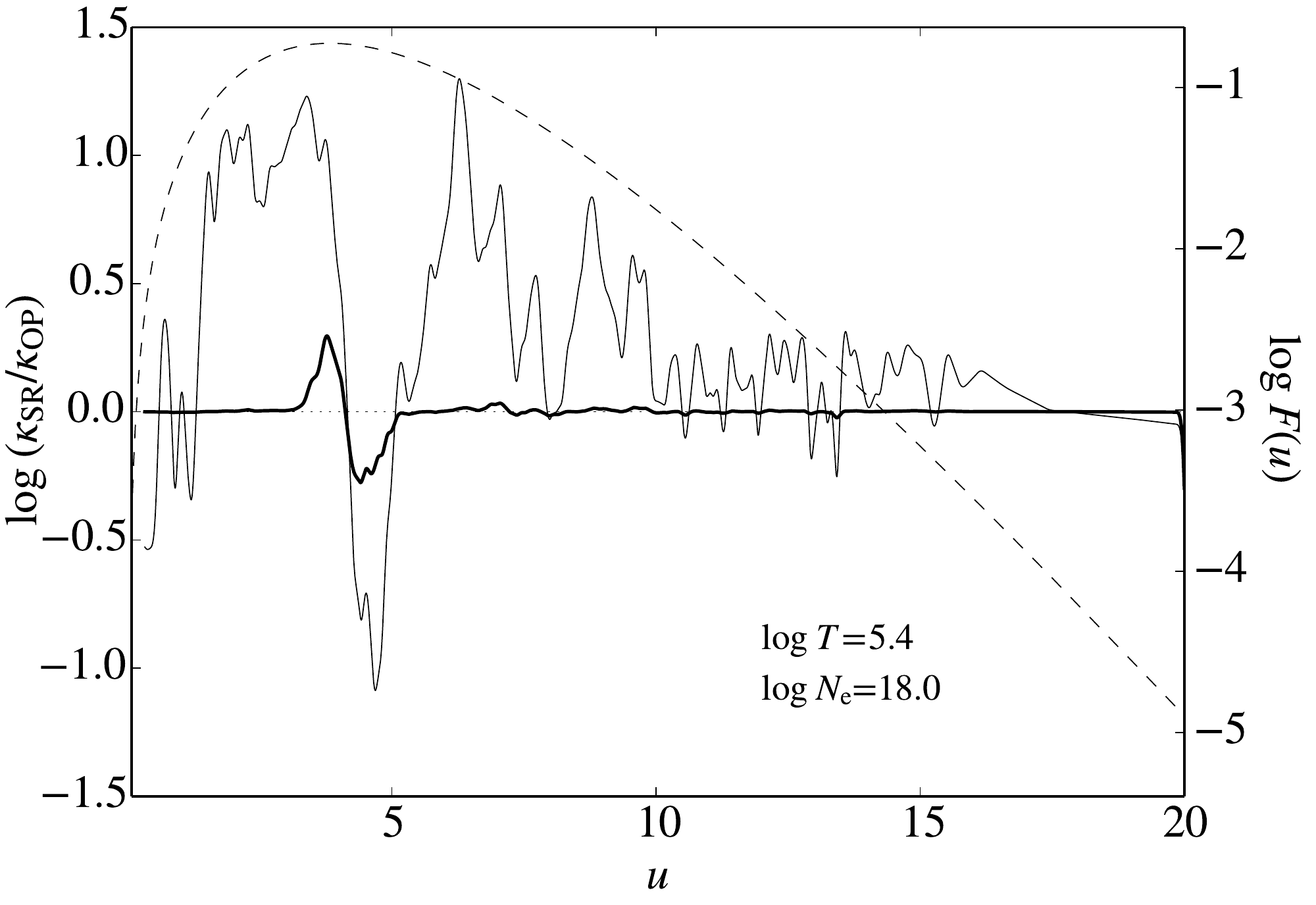}
      \caption{Logarithm of the ratio of the monochromatic cross sections for the stellar mixture with the new Ni data  ($\kappa_\mathrm{SR}$) and those with the OP Ni data ($\kappa_\mathrm{OP}$) vs. $u$, and the same quantity for Ni alone (thick and thin solid lines, respectively, left scale). The dashed line shows the $F(u)$ function (right scale). The dotted line represents a ratio of one. For the sake of legibility, the spectra have been convolved with a Gaussian kernel with a standard deviation of 0.05.
              }
         \label{figure:spectraRMO}
\end{figure}

This lack of correlation is due to the details of the cross-section spectrum of each chemical element of the stellar mixture. As an illustration, we detail the case in which the conditions in temperature and electron density yield a higher RMO for Ni alone, whereas that of the global mixture is reduced when the new data are used ($\log\ T$=5.4, $ \log\ N_\mathrm{e}$=18.0). Figure~\ref{figure:spectraRMO} shows the logarithm of the ratios of the cross sections of the global mixture with the new data and those with the OP values (thick line), and the same ratio for Ni alone (thin line). Plotting the logarithm of the ratios instead of the ratios themselves has the advantage of being meaningful in terms of cross sections, but also in terms of their inverses, as involved in the expression of the Rosseland mean, by considering the opposite. The ratio of the cross sections for the global mixture changes significantly in the interval $3<u<5.5$, whereas that of Ni departs from unity in the whole photon energy domain. We can therefore deduce that nickel contributes significantly to the cross sections of the global mixture only for $u$ in the interval [3, 5.5]. The ratio of the global cross section more or less follows that of Ni in this range. Around the maximum of the $F(u)$ function, the ratio for nickel is above unity over a wider range in $u$ than the one for which this ratio is below one, yielding an increase in the Ni Rosseland mean. In contrast, the interval for which the ratio of the global mixture is above one is smaller than the one where it is below one, along with deficiencies in opacity more important than their increase. The global RMO is thus smaller with the new data.

\begin{figure}
   \centering
   \includegraphics[width=.5\textwidth]{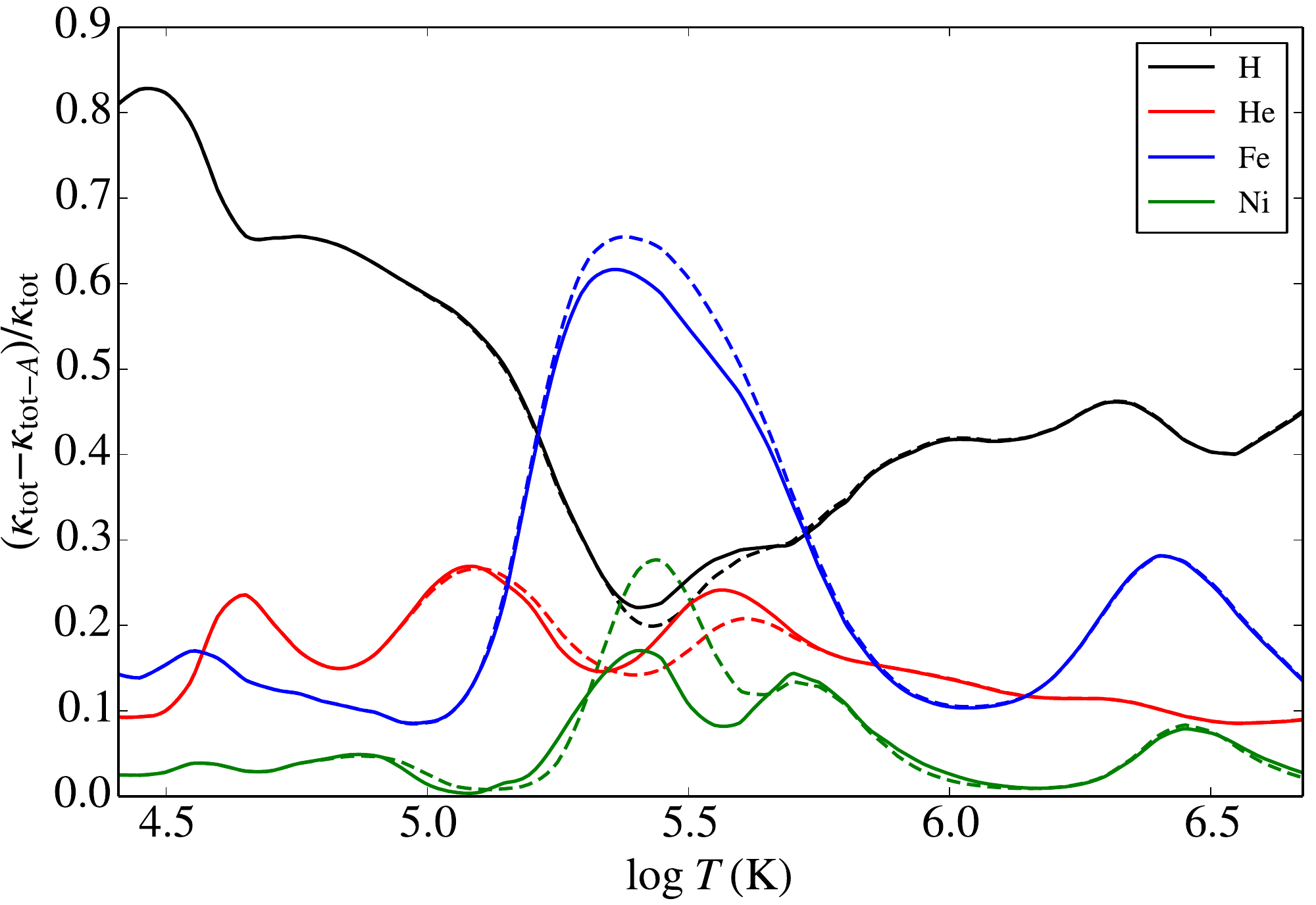}
      \caption{Contributions to the Rosseland mean opacities around the Z-bump as a function of $\log\ T$. Only the four most important contributors in these layers (H, He, Fe, and Ni) are represented. The solid lines correspond to the new data, and the dashed lines show the OP values. 
              }
         \label{figure:contribution}
\end{figure}

Since the shape of the Z-bump is modified with the new data, we expect that the contribution of the various elements to the RMO changes. The definition of this contribution is similar to that of the different absorption processes mentioned in Sect.~\ref{RMO}: for a given chemical element $A$, its contribution writes $(\kappa_{\mathrm {tot}}-\kappa_{\mathrm {tot}-A})/\kappa_{\mathrm {tot}}$ , where $\kappa_{\mathrm {tot}}$ is the Rosseland mean of the whole stellar mixture and $\kappa_{\mathrm {tot}-A}$ is the RMO without element $A$. Figure~\ref{figure:contribution} focuses around the Z-bump and includes the four most important contributors there, namely H, He, Fe, and Ni. With the new data, a change is expected for Ni, whose location of its maximum contribution is displaced at $\log\ T=5.4$ instead of $\log\ T=5.46$. In addition, its contribution is reduced compared to the OP data. The other elements are also affected by the new Ni cross sections because of the change in the Ni spectrum. With the new data, the interval in which the nickel opacity is strong around the maximum of the $F$ function overlaps more strongly in terms of photons energy with the interval in which iron has strong absorption features as well. That the photons are shared with iron in this energy range explains the reduced nickel contribution. For the same reason, the contribution of Fe is slightly reduced. The contributions of H and He, whose cross sections do not have any strong dependence versus photon energy near the maximum of the $F$ function, are enhanced because of the decrease of those of some other elements, mainly Fe.   

\begin{figure}
   \centering
   \includegraphics[width=.5\textwidth]{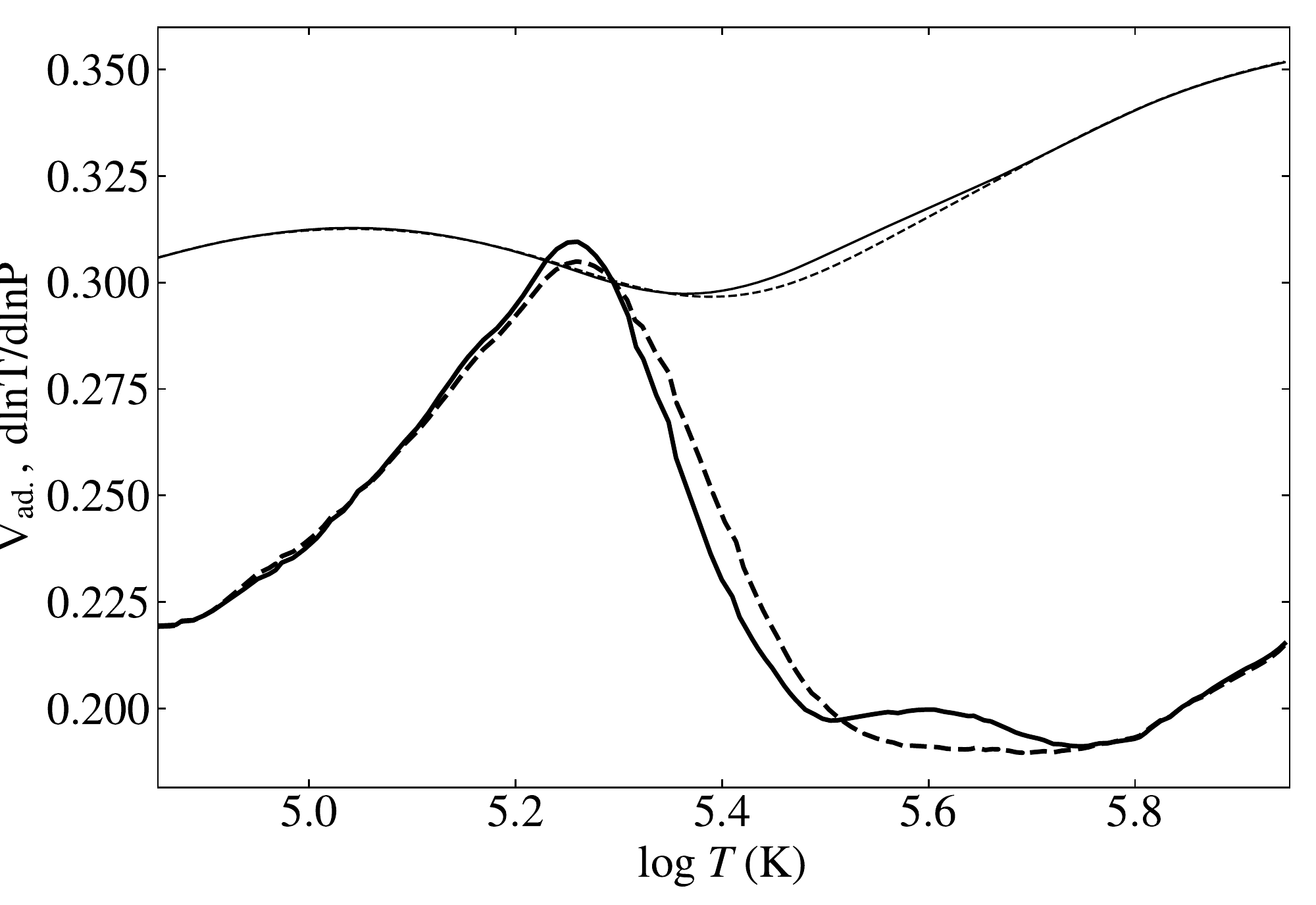}
      \caption{
       Adiabatic (thin lines), and real gradients (thick lines) as a function of $\log\ T$. The solid and dashed lines correspond to a model computed with the new Ni data and to a model calculated with Ni cross sections from OP, respectively.
              }
         \label{figure:compareGradients}
\end{figure}

Figure~\ref{figure:compareGradients} compares the adiabatic 
and real gradients for the models with the new data and the model with OP. The new data change the 
actual gradient in the whole Z-bump, whereas the adiabatic gradient is only affected around $\log\ T=5.6$.
The new data yield a slightly larger convective zone in the Z-bump compared to the OP model. The subsequent changes in the stellar structure and fundamental parameters are summarised in Table~\ref{table:structuralChanges}.

\begin{table}
   \centering
   \begin{tabular}{lc}
   \hline
   Quantity&Relative change\\
   \hline
   \hline
   $T_\mathrm{eff}$&$1.1\ 10^{-4}$\\
   $R_*$&$-2.1\ 10^{-4}$\\
   $\log g$&$4.2\ 10^{-4}$\\
   $T_\mathrm{C}$&$-4.3\ 10^{-5}$\\
   $\rho_\mathrm{C}$&$-4.4\ 10^{-5}$\\
   \hline
    \end{tabular}
      \caption{Relative changes in the stellar structure and fundamental parameters between the SCO-RCG model and the OP model. $T_\mathrm{C}$ and $\rho_\mathrm{C}$ are the central temperature and density, respectively. The other symbols have their usual meaning.
              }
         \label{table:structuralChanges}
\end{table}

\section{Discussion and conclusions}\label{discussion}

\begin{figure}
   \centering
   \includegraphics[width=.5\textwidth]{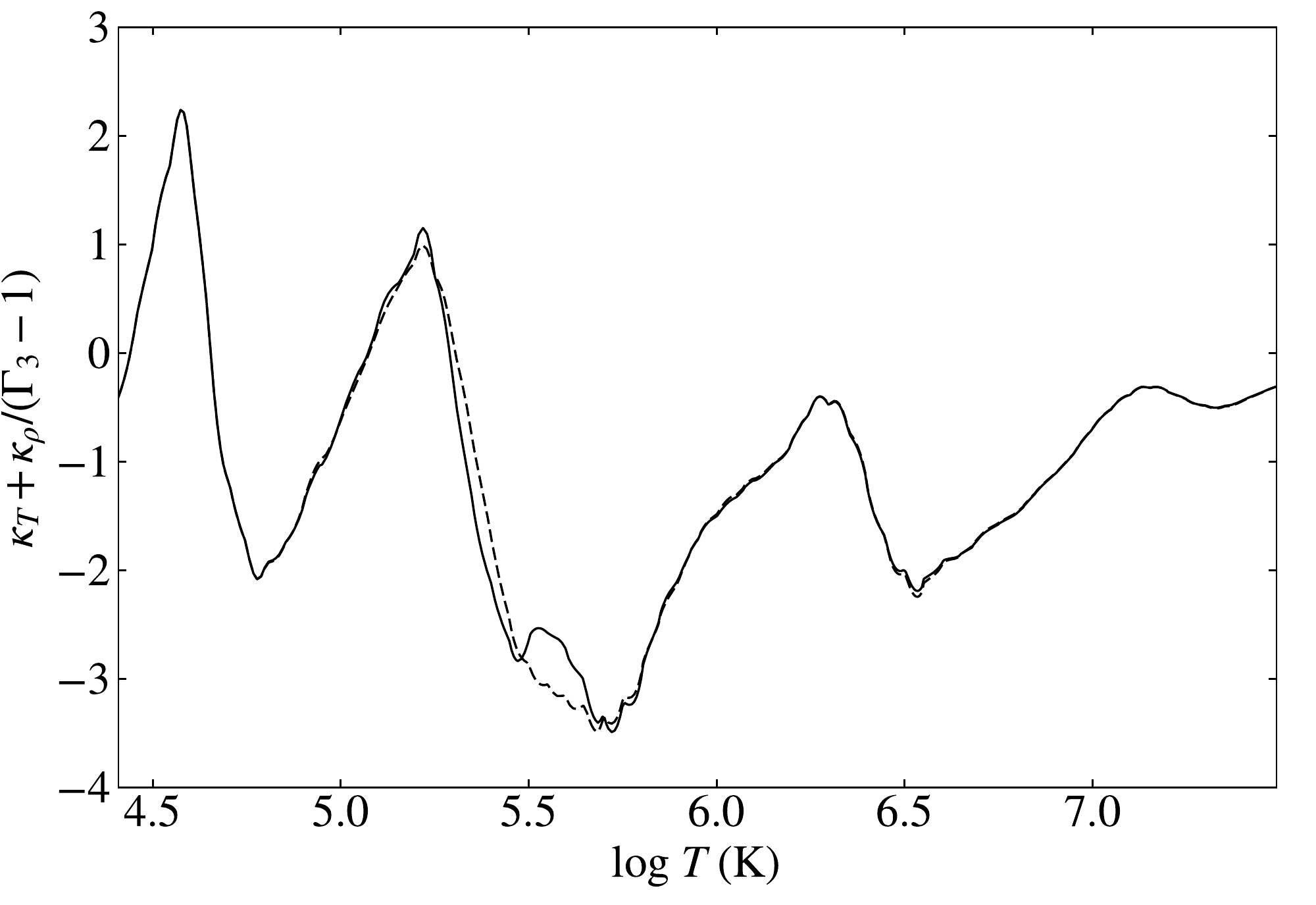}
      \caption{Temperature derivative of the Rosseland opacity vs. temperature. The solid line corresponds to the new data, and the dashed line shows the OP computation. 
              }
         \label{figure:derKappa}
\end{figure}

With Ni cross sections computed with SCO-RCG, the shape of the RMO versus $T$ in the Z-bump is only slightly altered. It has a narrower peak and a shallower maximum in temperature. The value of the maximum is almost unchanged. The temperature derivative of the Rosseland opacity is plotted in Fig.~\ref{figure:derKappa}. This derivative is related to the driving of pulsation modes \citep{Pamyatnykh1999} and reads $\kappa_T+\kappa_\rho /(\Gamma_3-1),$ where $\kappa_T=\frac{\partial\ln\kappa}{\partial\ln T}|_\rho$, $\kappa_\rho=\frac{\partial\ln\kappa}{\partial\ln \rho}|_T$, and $\Gamma_3-1=\frac{\partial\ln T}{\partial\ln\rho}|_{\mathrm {ad.}}$. It shows an additional maximum around $\log\ T=5.5$ with the new data, and the stability of certain pulsation modes are likely to be affected. The shape of this derivative is close to that obtained with OPAL or OPLIB by \cite{Walczak_etal2015} for their $10~M_\odot$ model. The details of a stability analysis are beyond the scope of the present paper, but we can already state that the new data for Ni are not able to reconcile the models with the observed pulsation modes of main-sequence massive pulsators because the RMOs we obtain in the Z-bump are too small compared to the values needed to reproduce individual objects \citep[][and references therein]{Walczak_etal2019} as well as the instability strips \citep{Moravveji2016}. This missing opacity could be produced by an accumulation of iron-peak elements, as suggested by \cite{Pamyatnykh_etal2004} and computed with self-consistent models by \cite{Hui-Bon-Hoa_Vauclair2018b,Hui-Bon-Hoa_Vauclair2018}. Forthcoming studies are planned to use the new Ni data to compute radiative accelerations and revisit their results. A stability analysis would then be worth performing to confirm the agreement of observed modes and those obtained with these models.

\begin{acknowledgements}
This work was supported by the "Programme National de Physique Stellaire" (PNPS) of CNRS/INSU co-funded by CEA and CNES.
\end{acknowledgements}

%
%

   \bibliographystyle{aa} 
   \bibliography{NiRMO}

\end{document}